\documentclass{aa}
\usepackage{graphicx, natbib,epsfig,amsmath,amssymb, mathrsfs, txfonts}
\DeclareGraphicsExtensions{.eps, .jpg, .ps, .png}
\bibpunct{(}{)}{;}{a}{}{,}
\begin{document}

\title{Design, analysis, and testing of a microdot apodizer for the Apodized Pupil Lyot Coronagraph. II. The dot size impact. \textit{(Research note)}}
\author{P. Martinez\inst{1} \and C. Dorrer\inst{2}  \and M. Kasper\inst{1} \and A. Boccaletti\inst{3} \and K. Dohlen\inst{4}}
\institute{European Southern Observatory, Karl-Schwarzschild-Strasse 2, D-85748, Garching, Germany 
\and Aktiwave, 241 Ashley drive, Rochester, NY, 14620-USA
\and LESIA, Observatoire de Paris Meudon, 5 pl. J. Janssen, 92195 Meudon, France
\and LAM, Laboratoire d\'{}Astrophysique de Marseille, 38 rue Fr\'{e}d\'{e}ric Joliot Curie, 13388 Marseille cedex 13, France}
\offprints{P. Martinez, martinez@eso.org}

\abstract
{The Apodized Pupil Lyot Coronagraph (APLC) is a promising coronagraphic device for direct exoplanets detection on the European-Extremely Large Telescope.
This concept features amplitude apodization in the entrance aperture, and a small opaque Lyot mask in the focal plane.  
We present new near-IR laboratory results using binary apodizers -- the so-called microdots apodizer -- which represent a very attractive and advantageous solution for the APLC.}
{Microdots apodizers introduce high-frequency noise whose characteristics depend on the pixel size. The aim of this work is to characterize the impact of the pixel size on the coronagraphic image. 
Estimation of both the noise intensity and its localization in the field of view is the objective of this study.}
{Microdots apodizer, consisting of an array of pixels with spatially variable density, that are either opaque or transparent, were manufactured by lithography of a light-blocking metal layer deposited on a transparent substrate. A set of 5 masks has been designed with different pixel sizes, tested in the near-IR, and their behavior compared to theoretical models.}
{Stray light diffraction introduced by the finite pixel size was measured during experiment. Intensity decreases, and radial distance increases, when the pixel size gets smaller.}
{The physical properties of these microdots apodizers have been demonstrated in laboratory. The microdots apodizer is a suitable solution for any coronagraph using pupil amplitude apodization if properly designed.}

\keywords{\footnotesize{Techniques: high angular resolution --Instrumentation: high angular resolution --Telescopes} \\} 

\maketitle


\section{Introduction}
Overcoming the large contrast between bright astrophysical sources and sub-stellar companions is mandatory for direct detection, and spectroscopy of extra-solar planets. 
In this context, a coronagraph used downstream of an XAO system is a powerful tool to improve the sensitivity of an imaging system to faint structures surrounding a bright source.
Efficient XAO systems are required to correct wavefront errors due to the atmospheric turbulence, while coronagraphs are designed to suppress, or at least attenuate the starlight diffracted by the telescope. 

The Apodized Pupil Lyot Coronagraph \citep[APLC]{2002A&A...389..334A, 2003A&A...397.1161S} is a promising design for any forthcoming planet-finder instruments on large ground-based telescopes \citep{2005ApJ...618L.161S, 2007A&A...474..671M, Corono, 2009ApJ...695..695S}. 
The main technical challenge in building an APLC is the apodizer.
Prototypes are currently being developed using different approaches and tested to validate the concept in laboratory conditions \citep{Reynard, HOT, APLCtest, microdots}. 

In this paper, we present near-IR laboratory results to characterize a new technological solution for apodizer manufacture.
In a former study \citep[hereafter Paper I]{microdots}, we explored a halftone-dot process to generate an array of binary pixels, either transparent or opaque, where the density of opaque pixels is spatially varying. This technique aims to solve the drawbacks of a continuous deposit of a metal layer with spatially varying thickness. 
Main advantages of a microdots apodizer are:
(1/) accuracy of the transmission profile, (2/) achromaticity, (3/) no spatially-varying phase, (4/) reproducibility.\\
In Paper I, we reported that microdots apodizers exhibit blue noise properties (i.e. high spatial frequency noise), when designed for coronagraphy.
Although numerical simulations and theoretical predictions confirm noise in the coronagraphic image, its impact was found to be negligible during experiments. 
Our first prototype (mask 1, hereafter) was designed to push this noise out of our field of view of interest at a deep contrast level, as a result of a fine adjustment of the pixel size (4.5$\mu$m).
Here, our purpose is to investigate the noise properties using 5 new masks with the same transmission profile as mask 1, but by successively increasing the pixel size. 
The interest is twofold: (1/) confirm the theoretical predictions on the physical properties of such devices with laboratory experiments, (2/) derive general relevant information for the design of amplitude microdots apodization mask for coronagraphy (whatever the coronagraph).  
Section 2 briefly reviews microdots apodizer theoretical properties, while Sect. 3 describes the experiment as well as the manufacturing details of the new masks. In Sect. 4 we present and discuss results obtained in the experiment, and finally in Sect. 5 we draw conclusions. 
\begin{figure*}[!ht]
\begin{center}
\includegraphics[width=8.5cm]{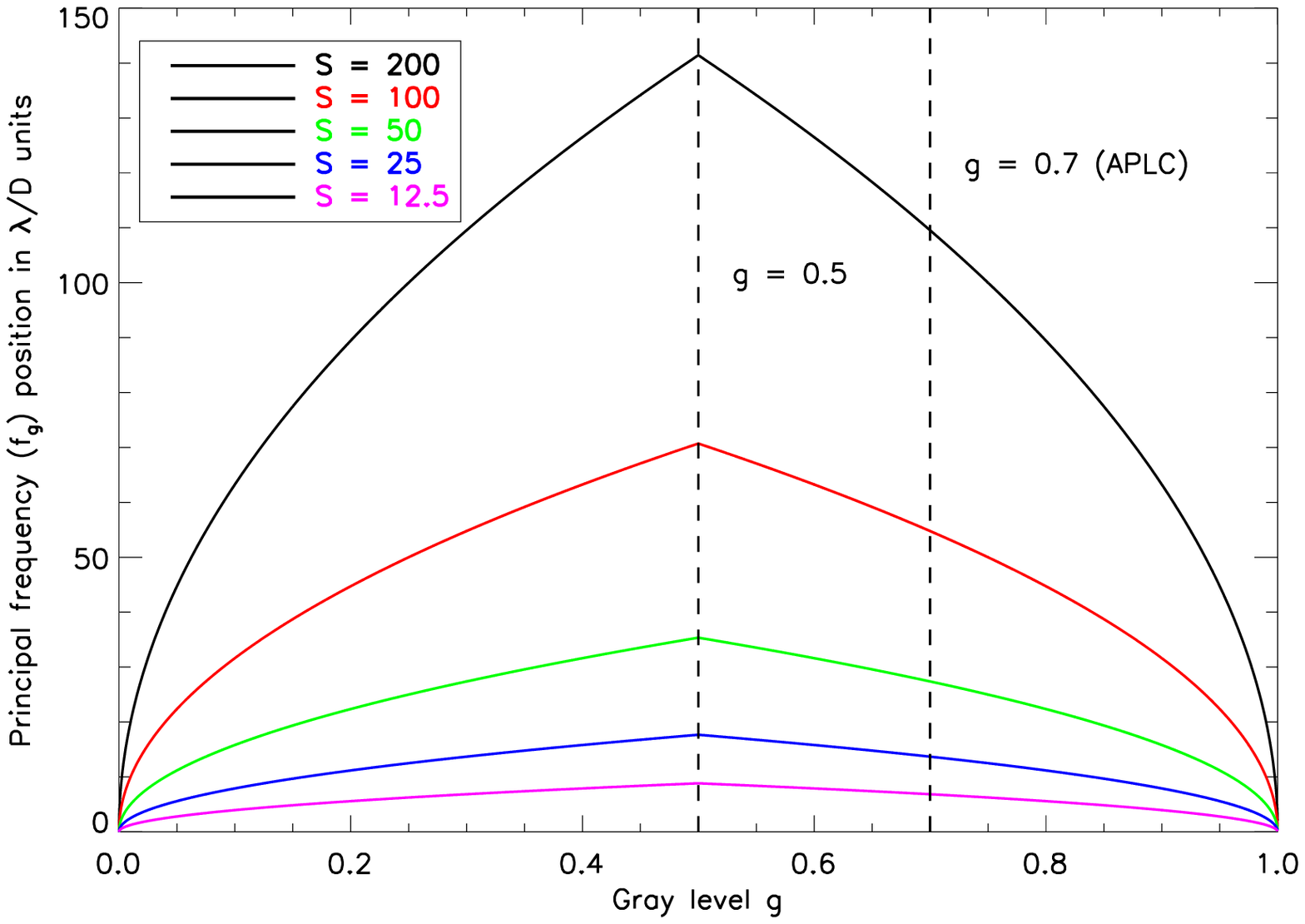}
\includegraphics[width=8.5cm]{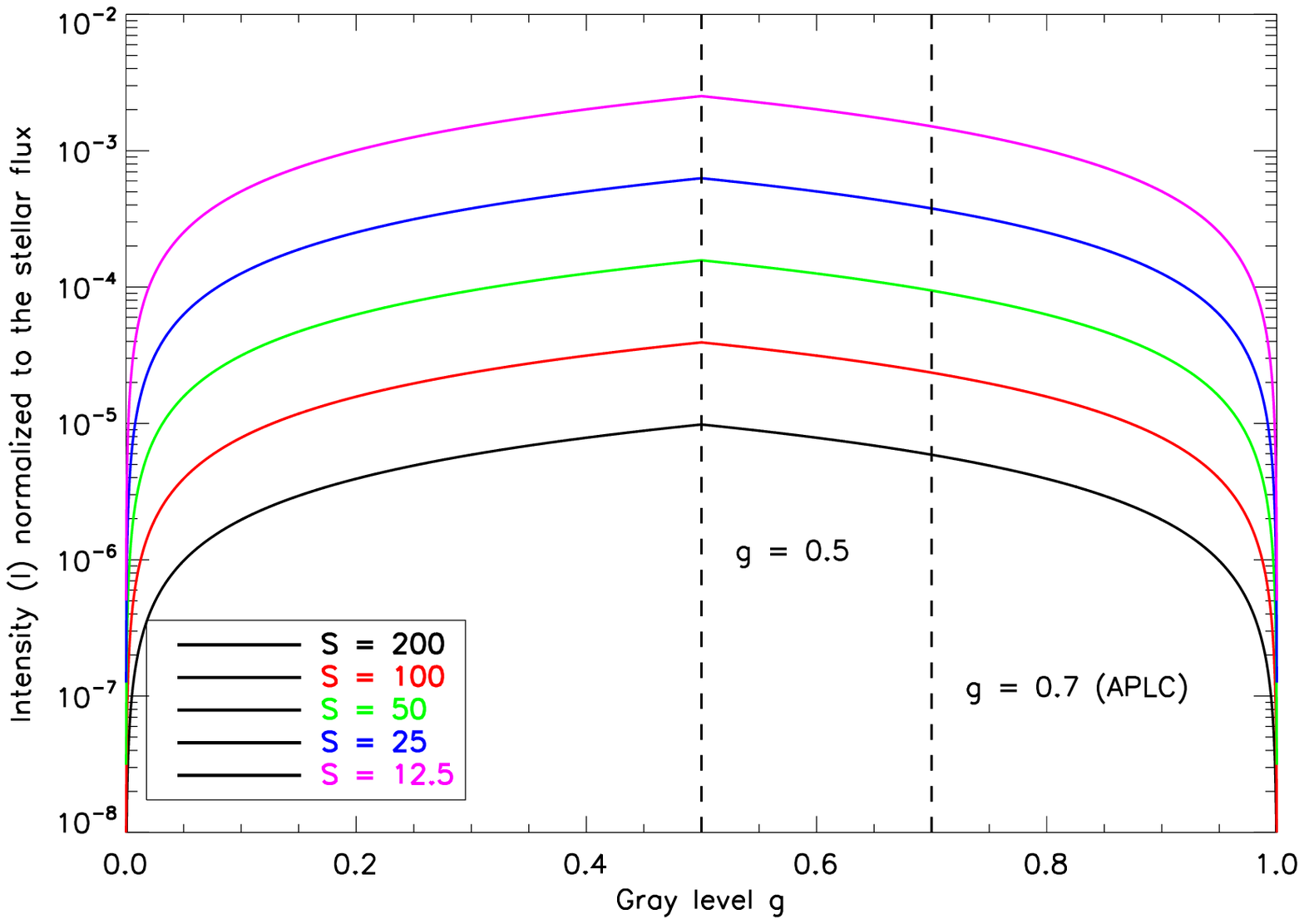}
\end{center}
\caption{$Left$: First order diffraction peak position ($f_g$ in $\lambda/D$ units) as a function of gray level $g$. $Right$: Speckles halo intensity $I$ normalized to the reference star intensity as function of gray level $g$. In both plots, the APLC case ($g=0.7$) is localized with dashed line.} 
\label{pixelsize}
\end{figure*} 

\section{Microdots apodizer theoretical properties}
For the reader's convenience, we briefly review the microdots apodizer theoretical properties (i.e. the signature of the dots in the coronagraphic image) using the same notation as defined in Paper I.
A microdots apodizer is modeled as an aperiodic under-filled two-dimensional grating. Such a device exhibits blue noise properties owing to the error diffusion algorithm used to calculate a distribution of pixels (i.e. dots) that best approximates the required field transmission \citep{floyd, Ulichney, 2007JOSAB..24.1268D}.
The binary pattern produces an average gray level value ($g = \sqrt{T}$, i.e average amplitude transmission) from an apodizer profile with intensity transmission $T$. 
\noindent The resulting noise spectrum of such a device is set by the minority pixels present on the device (i.e. by $g$, non-metal pixels when $g < 0.5$ and by metal pixels conversely).
The spectral energy therefore increases as the number of minority pixels increases, peaking at $g=0.5$ \citep{Ulichney}.
In the precise case of square pixels \citep{Ulichney2}, the noise spectrum of the pattern exhibits energy concentration around a first order diffraction peak ($f_g$) localized in the field of view in $\lambda/D$ units as:
\begin{equation}
f_g =
\begin{cases}
\sqrt{g}\times S  & g \leq 1/2 \\
\sqrt{(1-g)}\times S  & g > 1/2 \\
\end{cases}
\label{principalF}
\end{equation}
\noindent where S is the scaling (or sampling), ratio between the pupil diameter ($\Phi$) and the pixel size, i.e. dot size ($p$).
Higher order diffraction peaks are less relevant since out of the field of view when dots are small enough.
Diffraction order peaks are separated by $S$ in $\lambda/D$ units with $S$ extent, i.e. dots scatters light by diffraction and creates a 2D-sinus cardinal function halo in the focal plane.

In Paper I we presented a simplified model for order-of-magnitude estimation of the high-spatial frequency noise intensity in coronagraphic systems \citep{Dust}.
The speckle halo in the coronagraphic image resulting from the non-regular dots distribution broaden the first order diffraction peak $f_g$ with an intensity $I_g$ defined as:
\begin{equation}
I_g = 
\begin{cases}
g \times \frac{\pi}{4} \times \left(\frac{1}{S}\right)^{2} & g \leq 1/2 \\
(1-g) \times \frac{\pi}{4} \times \left(\frac{1}{S}\right)^{2} & g > 1/2 \\
\end{cases}
\label{Intensity}
\end{equation}
\noindent In Fig. \ref{pixelsize} we plotted the first order diffraction halo localization in the field, and its intensity (normalized by the stellar flux), as a function of the gray level for the set of scaling factors ($S$) we used for prototyping.
Decreasing the scaling factor, i.e. increasing the pixel size, therefore moves the first order diffraction halo closer to the central core of the coronagraphic image with an increase of intensity.  

\section{Experiment}
\subsection{Masks design and optical setup}
The configuration of the apodizer profile is similar to that described in Paper I (4.5 $\lambda/D$ APLC, $\Phi = 3mm$ due to constraints on our optical bench).
The 5 new apodizer masks were fabricated by Precision Optical Imaging in Rochester, New York. The amplitude pupil masks were fabricated using wet etch contact lithography of a regular Chrome layer ($OD = 4$) deposited on a BK7 glass substrate ($\lambda/20$ peak-to-valley), with antireflection coating for the H band (1.2 to 1.8$\mu$m, $R<1\%$) on their back faces.

Mask 1 had a specified scaling factor of 500 corresponding to 6$\mu$m pixels grid, but finally appeared being smaller (4.5$\mu$m), as a predictable result of the manufacturing process. 
Hereafter microdots mask 2, 3, 4, 5, and 6, have scaling factors specified to 200, 100, 50, 25, and 12.5 corresponding to 15, 30, 60, 120, and 240$\mu$m pixels respectively. Therefore, pixels size, i.e. dots size, increases by a factor of 2 from mask to mask. Table \ref{table1} gathers all the microdots masks characteristics and noise properties predicted by theory (Eq. \ref{principalF} and \ref{Intensity}).
The experimental configuration is similar to that described in Paper I. The optical setup is designed to simulate the 8-m VLT pupil and to operate in the near-infrared (the H-band). The Strehl ratio of the bench is $\sim94\%$.
The IR camera used (the Infrared Test Camera) uses a HAWAII $1k\times1k$ detector. Experiment was done in H-band using either a narrow filter ($\Delta \lambda/\lambda =1.4\%$) or a broadband filter ($\Delta \lambda/\lambda = 20\%$). 
The APLC pupil-stop is also similar to that of paper I and remains the same during the experiment. 
\begin{figure*}[!ht]
\begin{center}
\includegraphics[width=5cm]{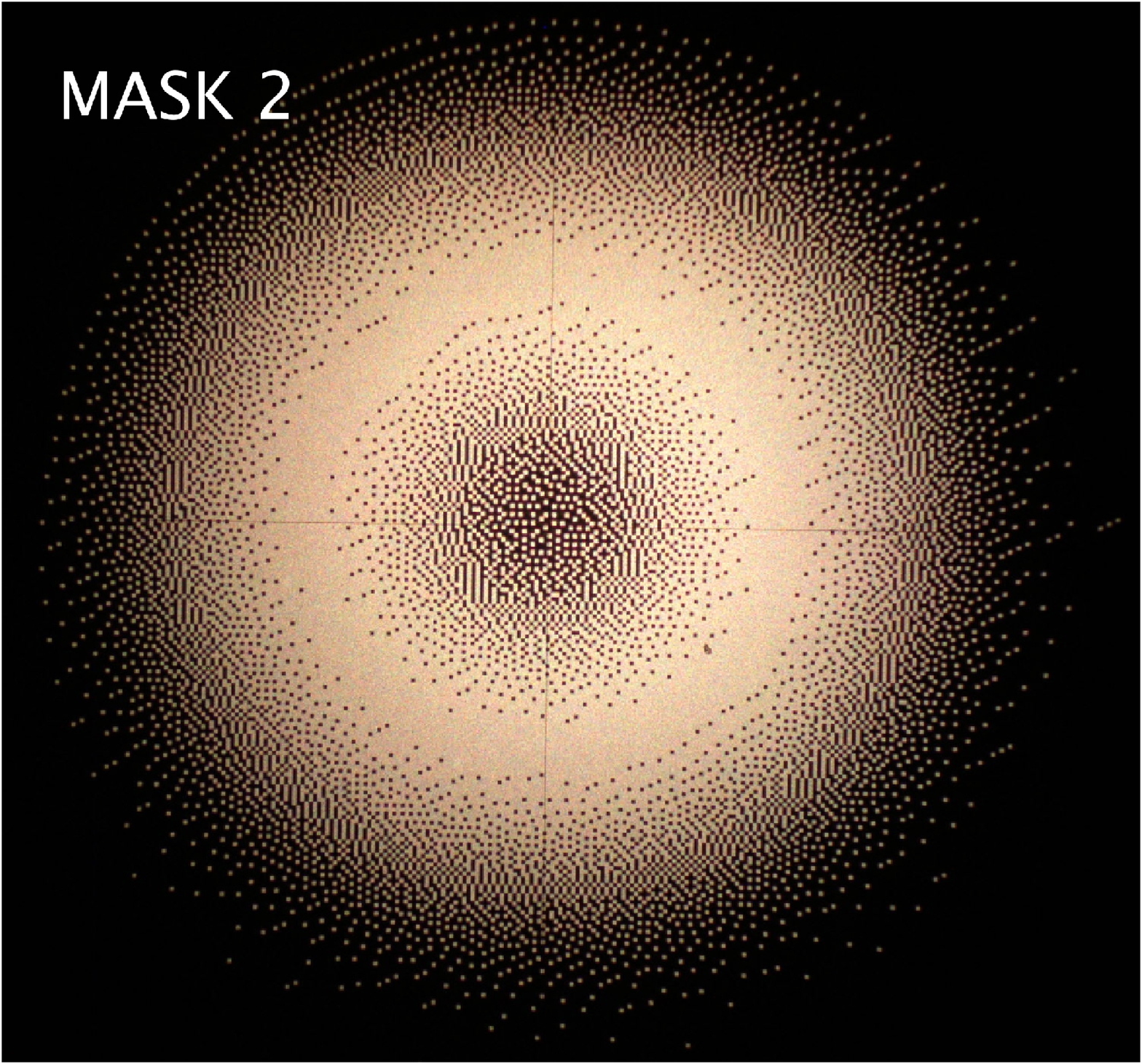}
\includegraphics[width=4.6cm]{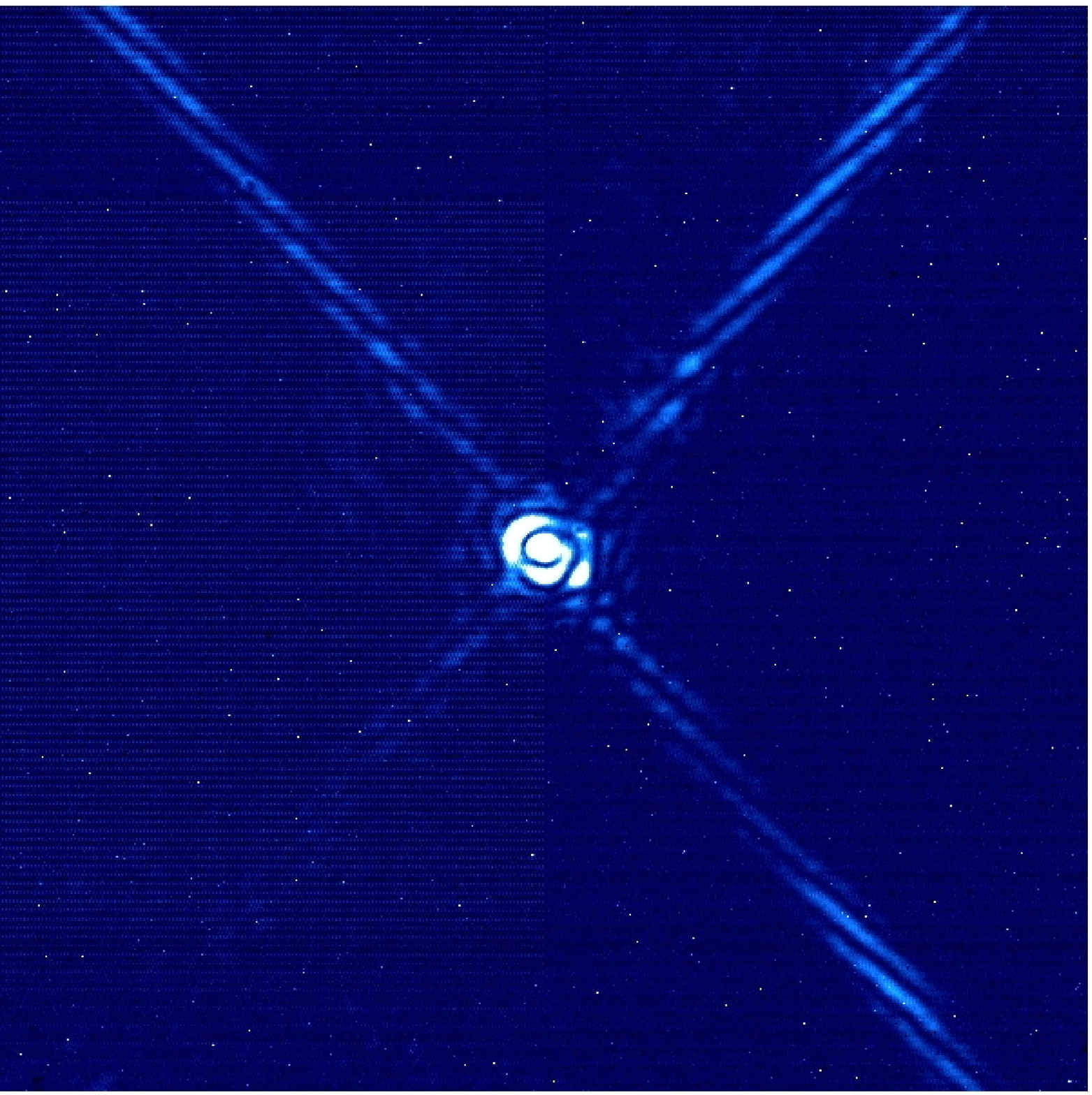}
\includegraphics[width=4.6cm]{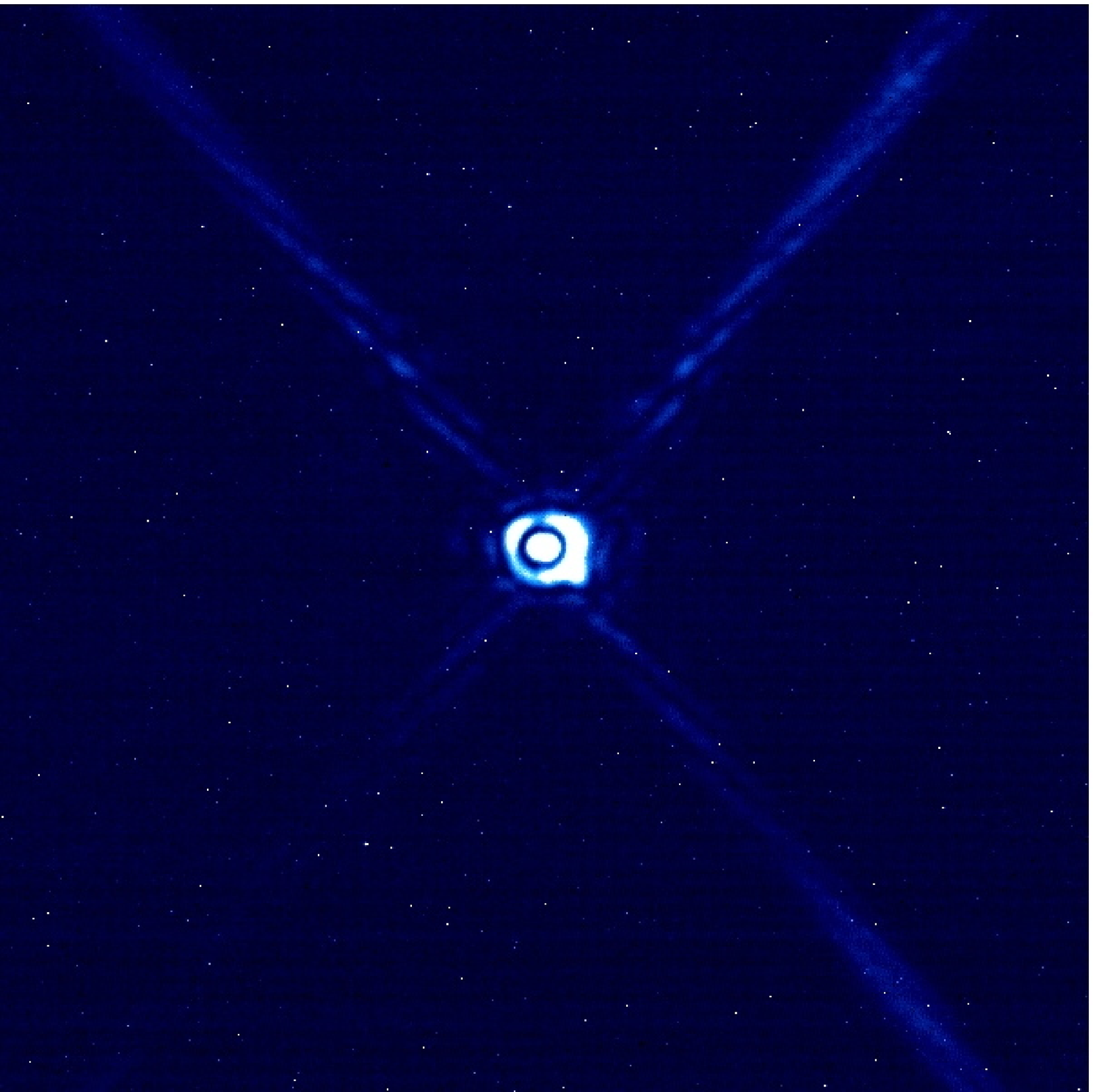}

\includegraphics[width=5cm]{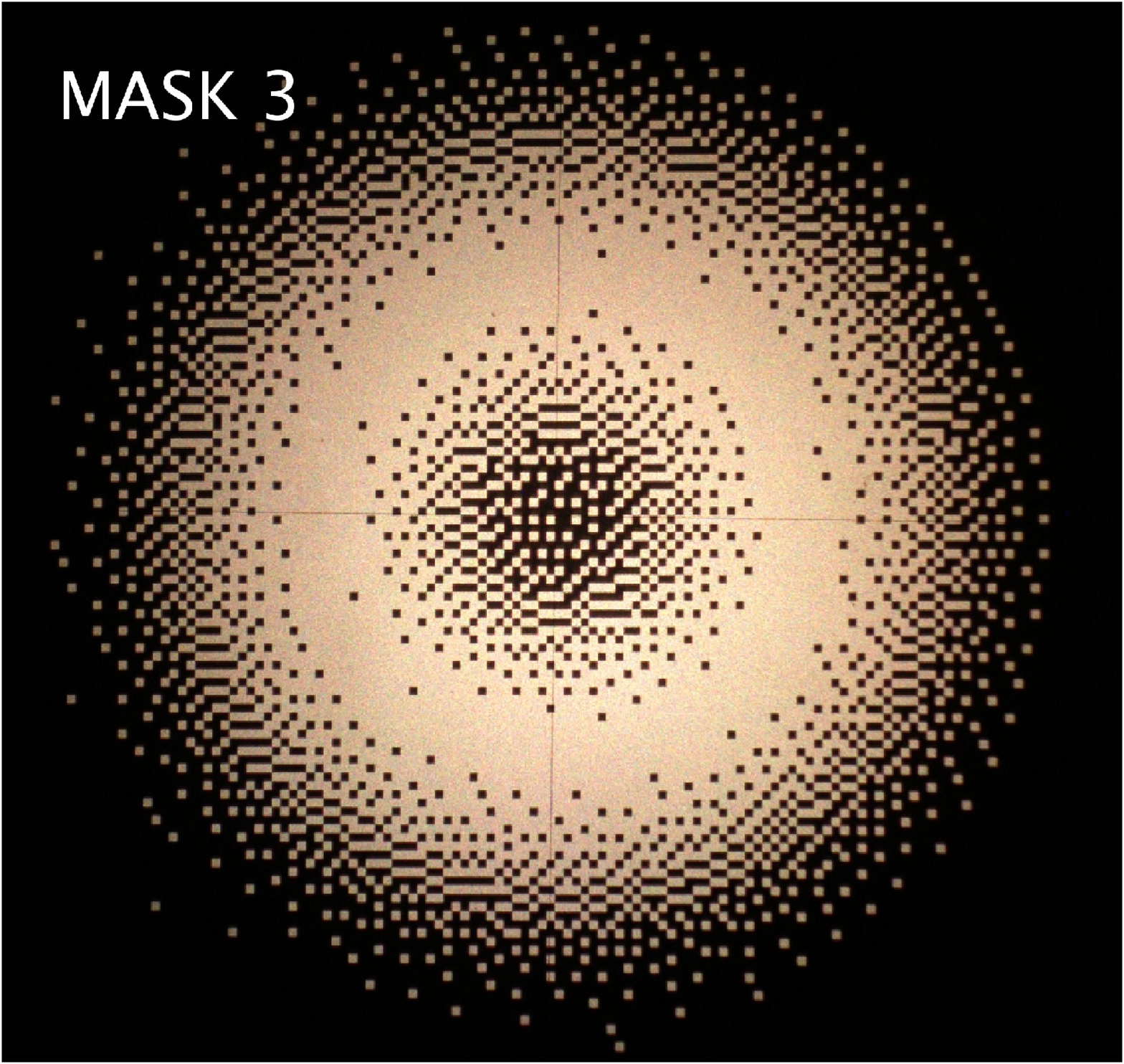}
\includegraphics[width=4.6cm]{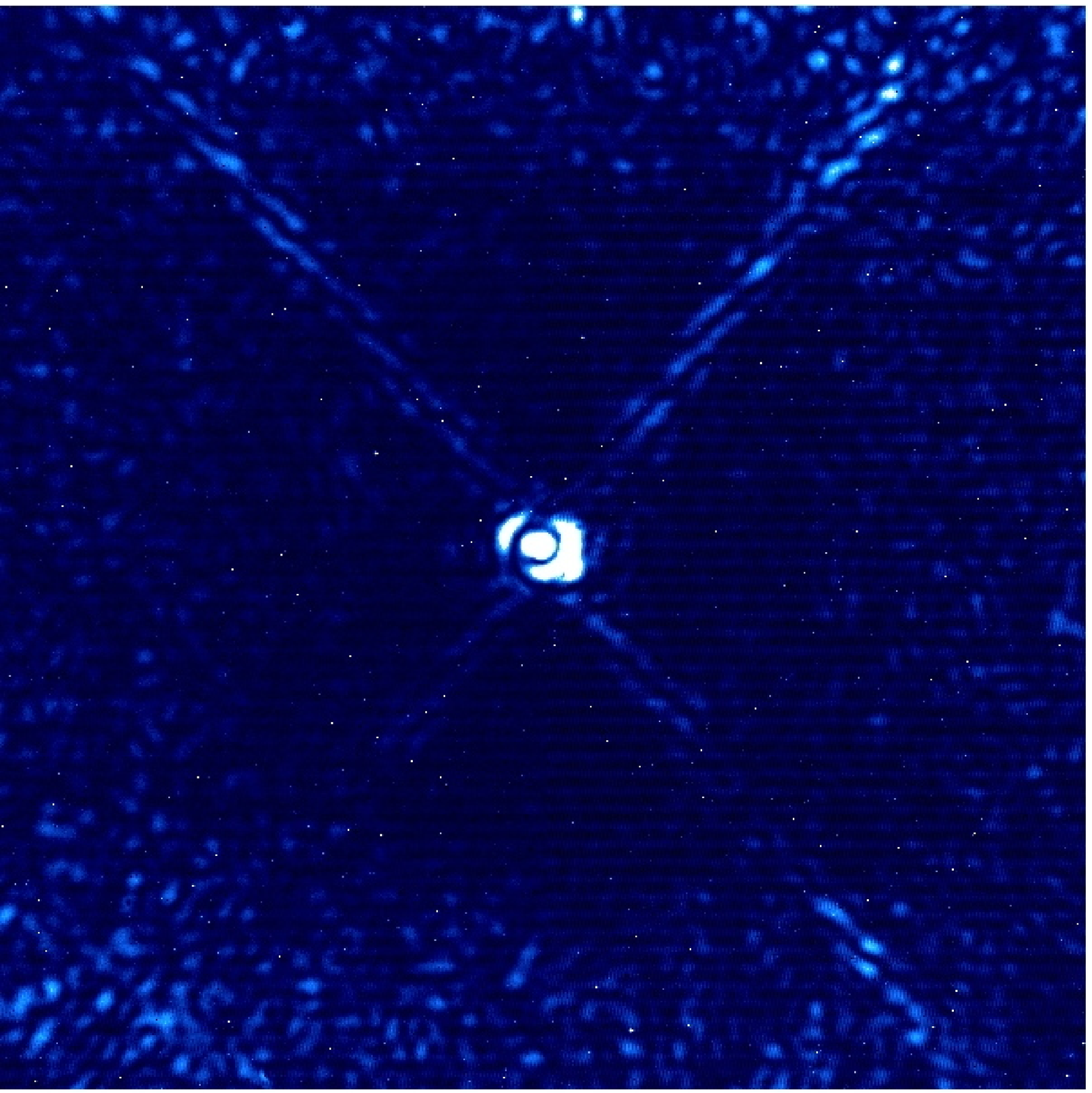}
\includegraphics[width=4.6cm]{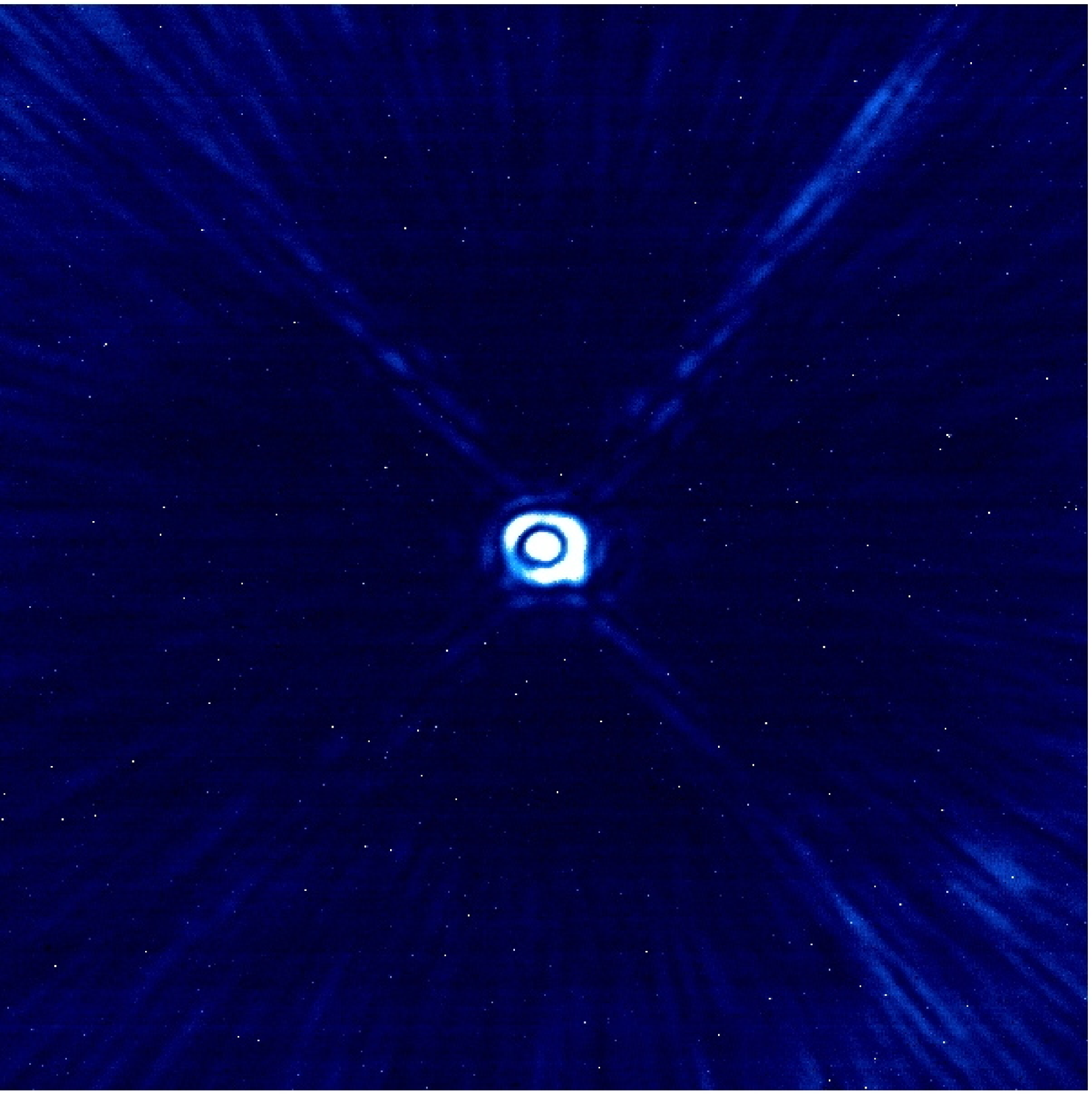}

\includegraphics[width=5cm]{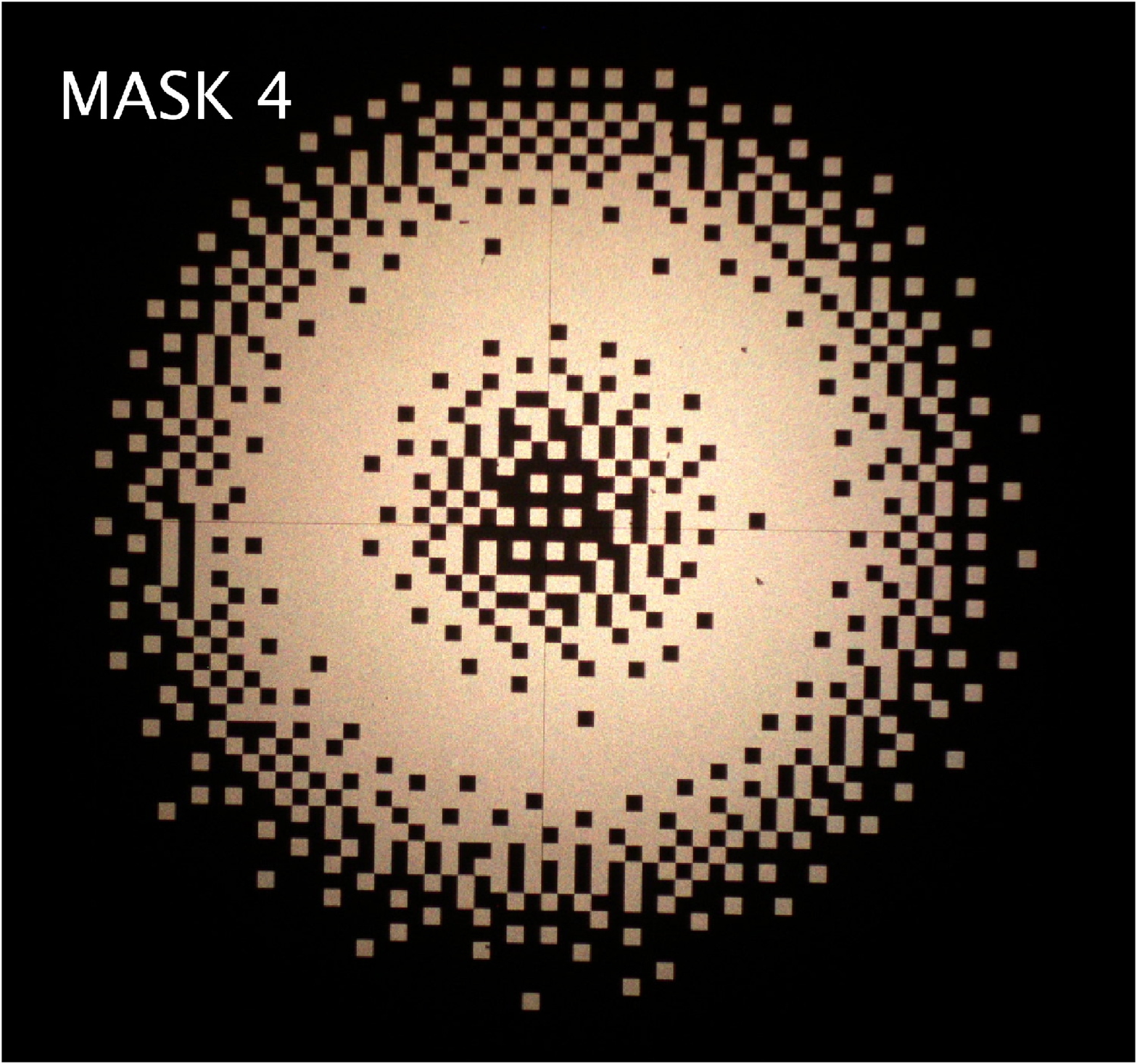}
\includegraphics[width=4.6cm]{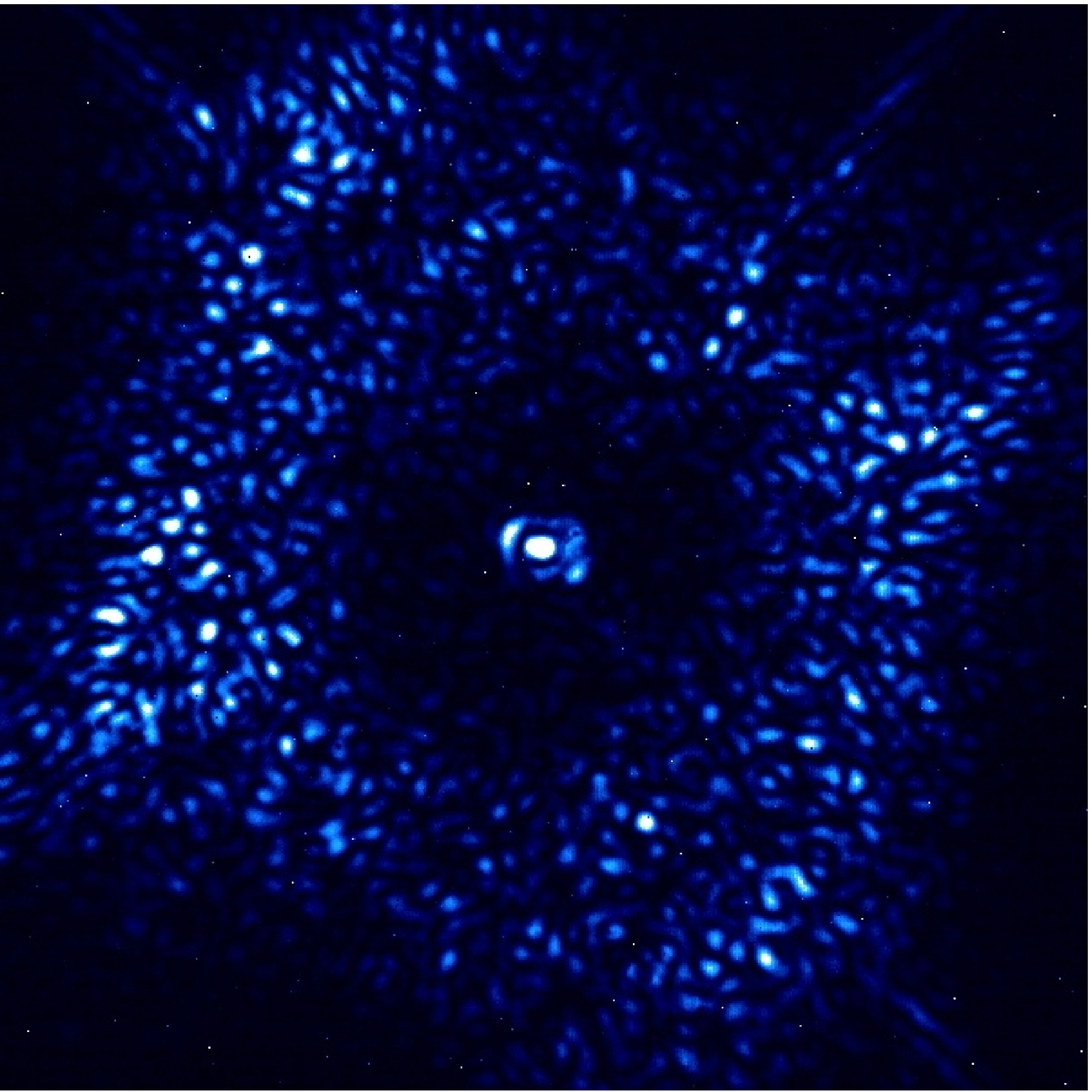}
\includegraphics[width=4.6cm]{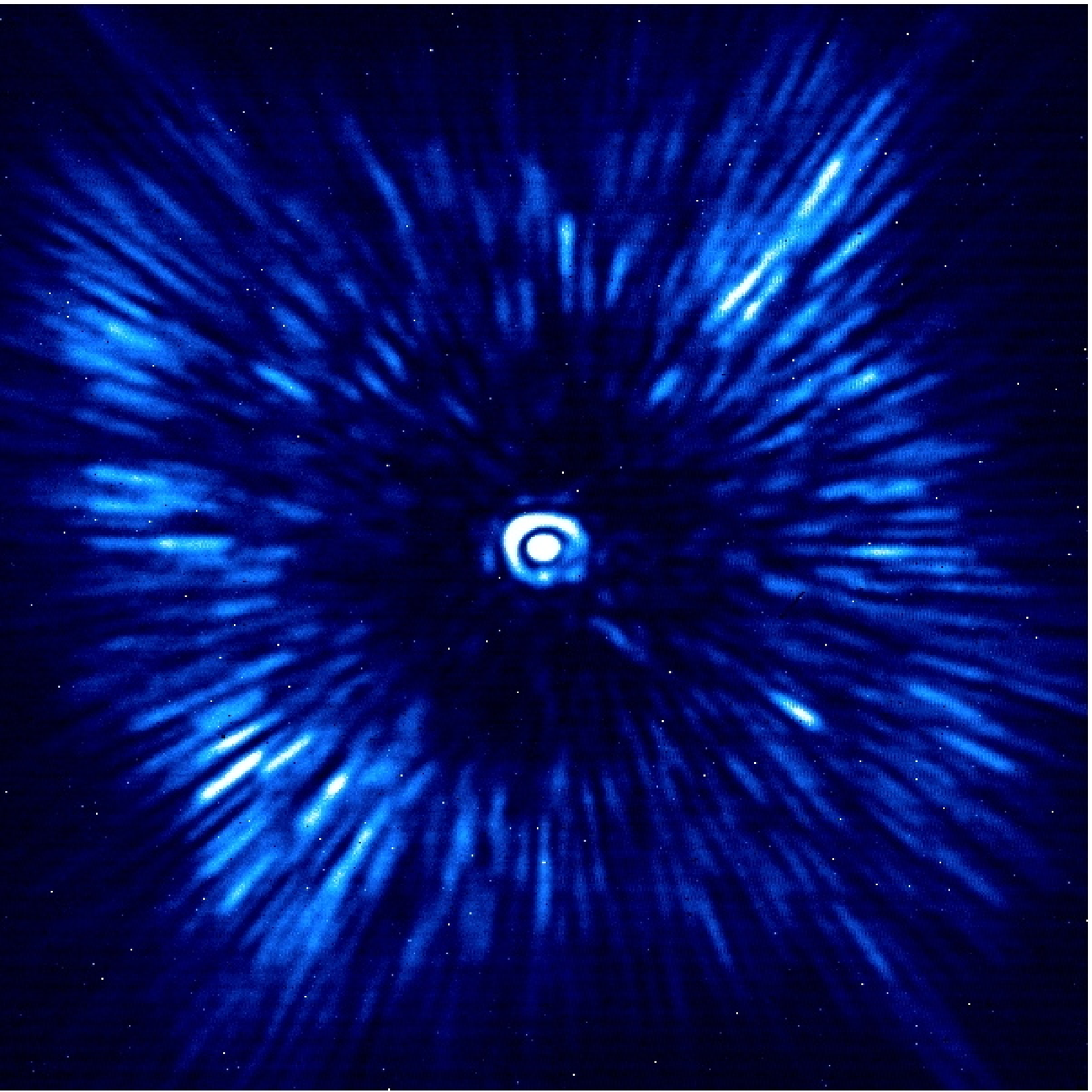}

\includegraphics[width=5cm]{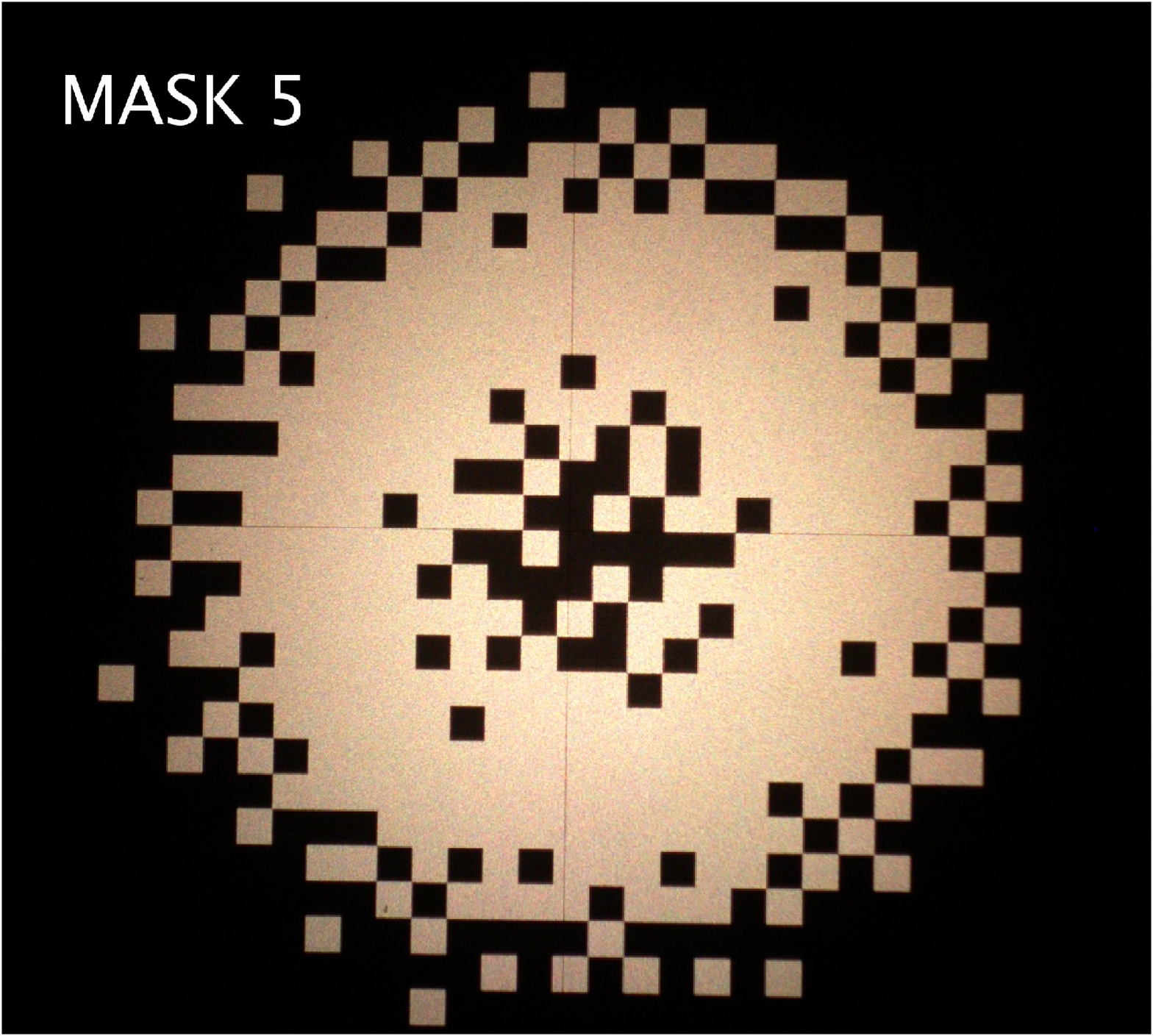}
\includegraphics[width=4.6cm]{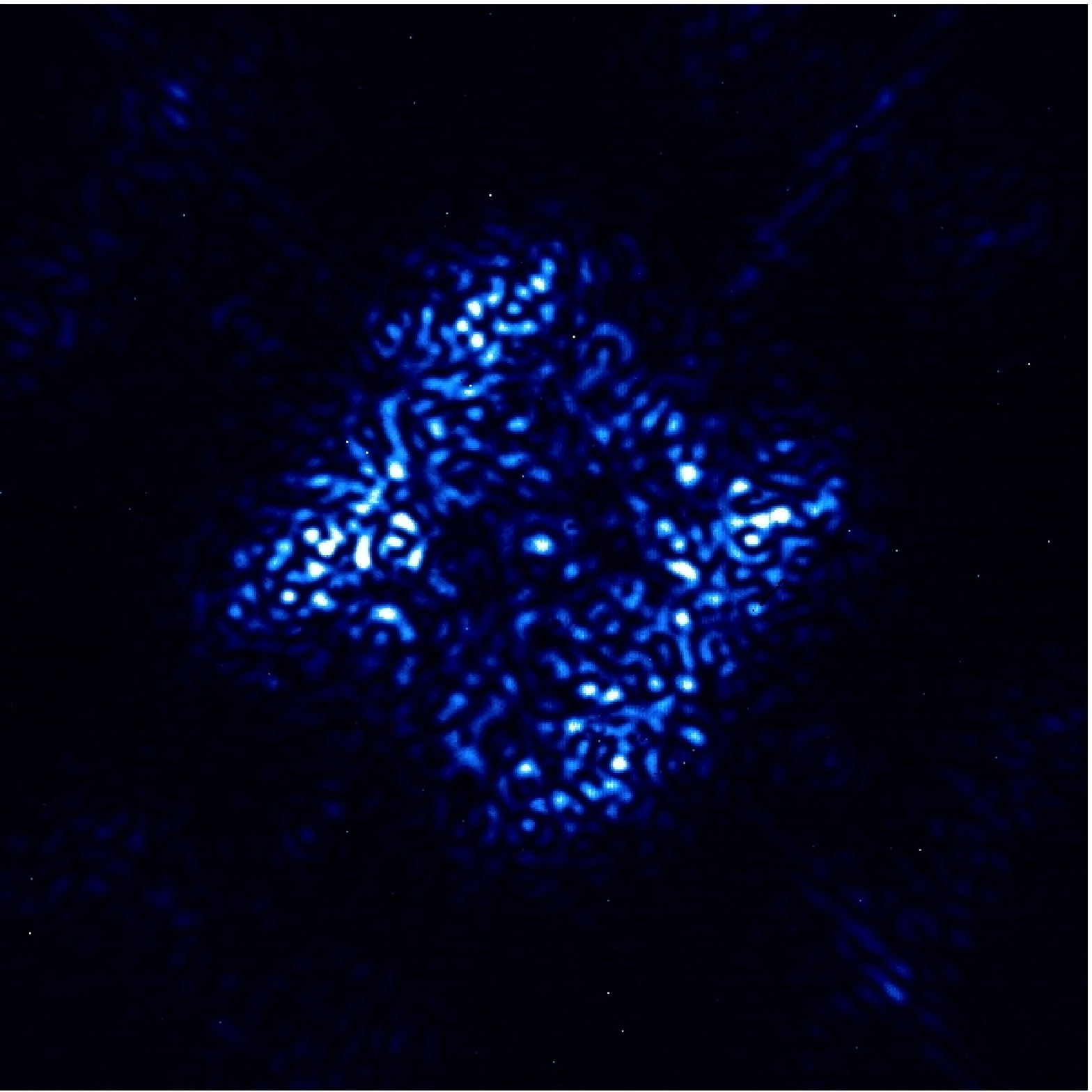}
\includegraphics[width=4.6cm]{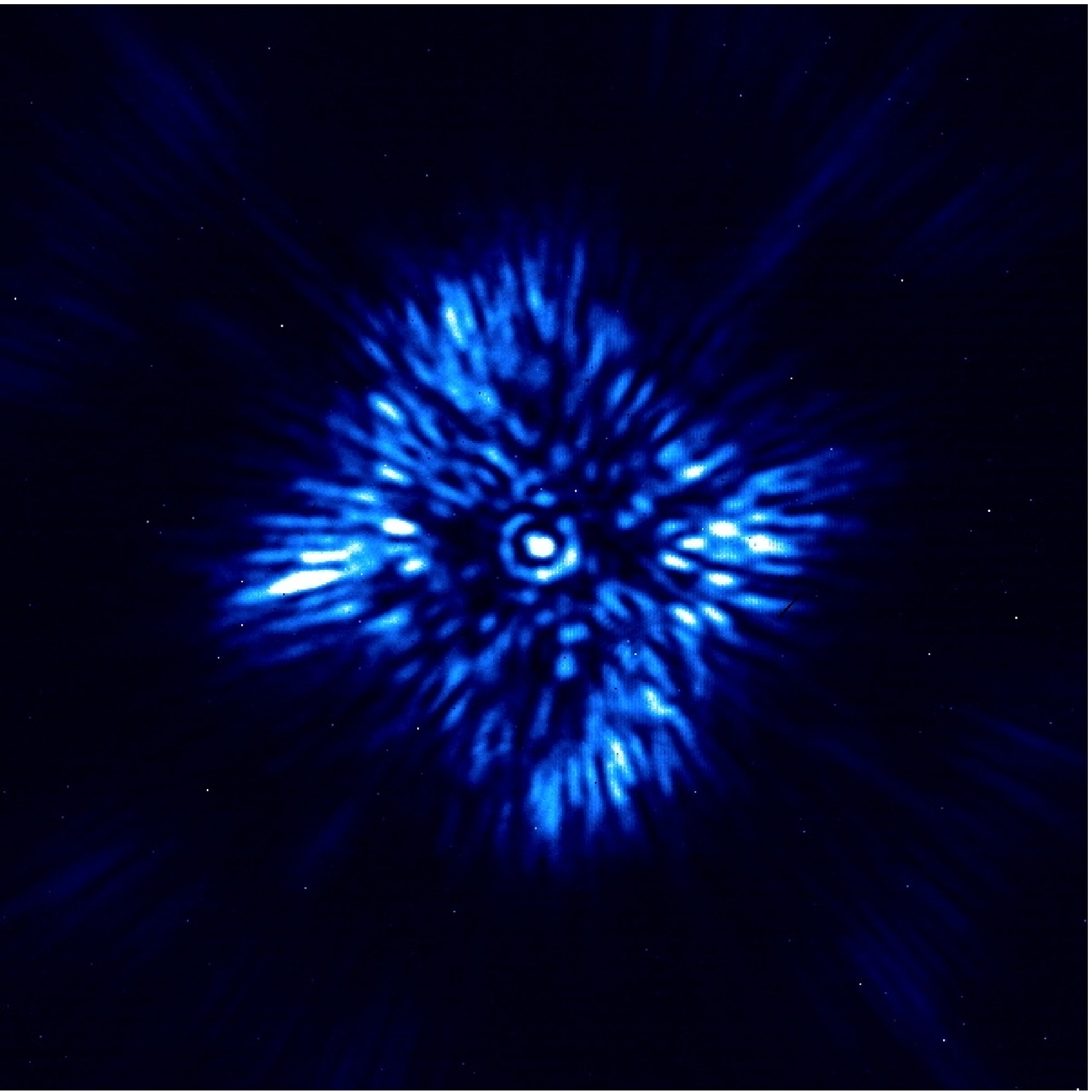}

\includegraphics[width=5cm]{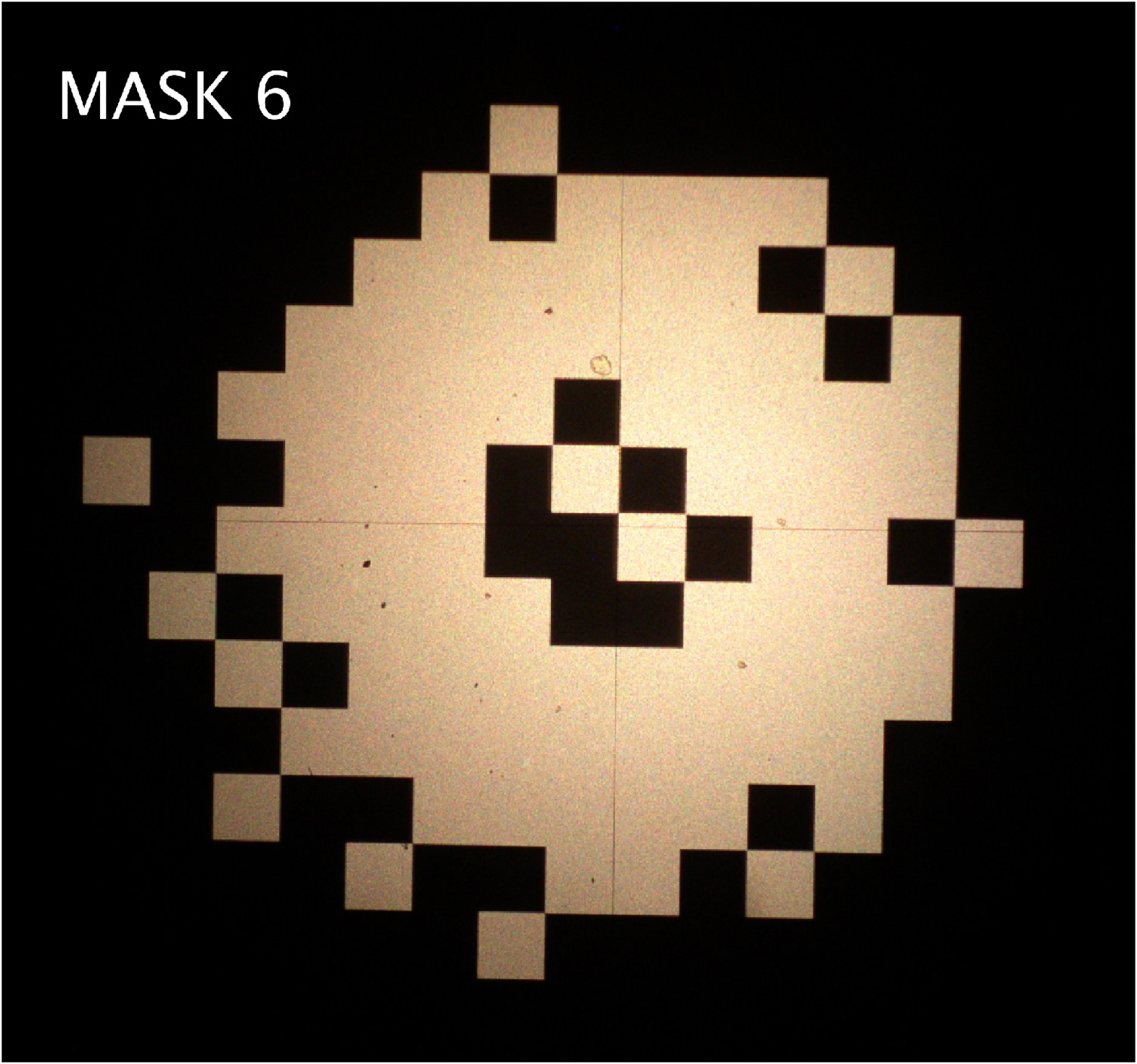}
\includegraphics[width=4.6cm]{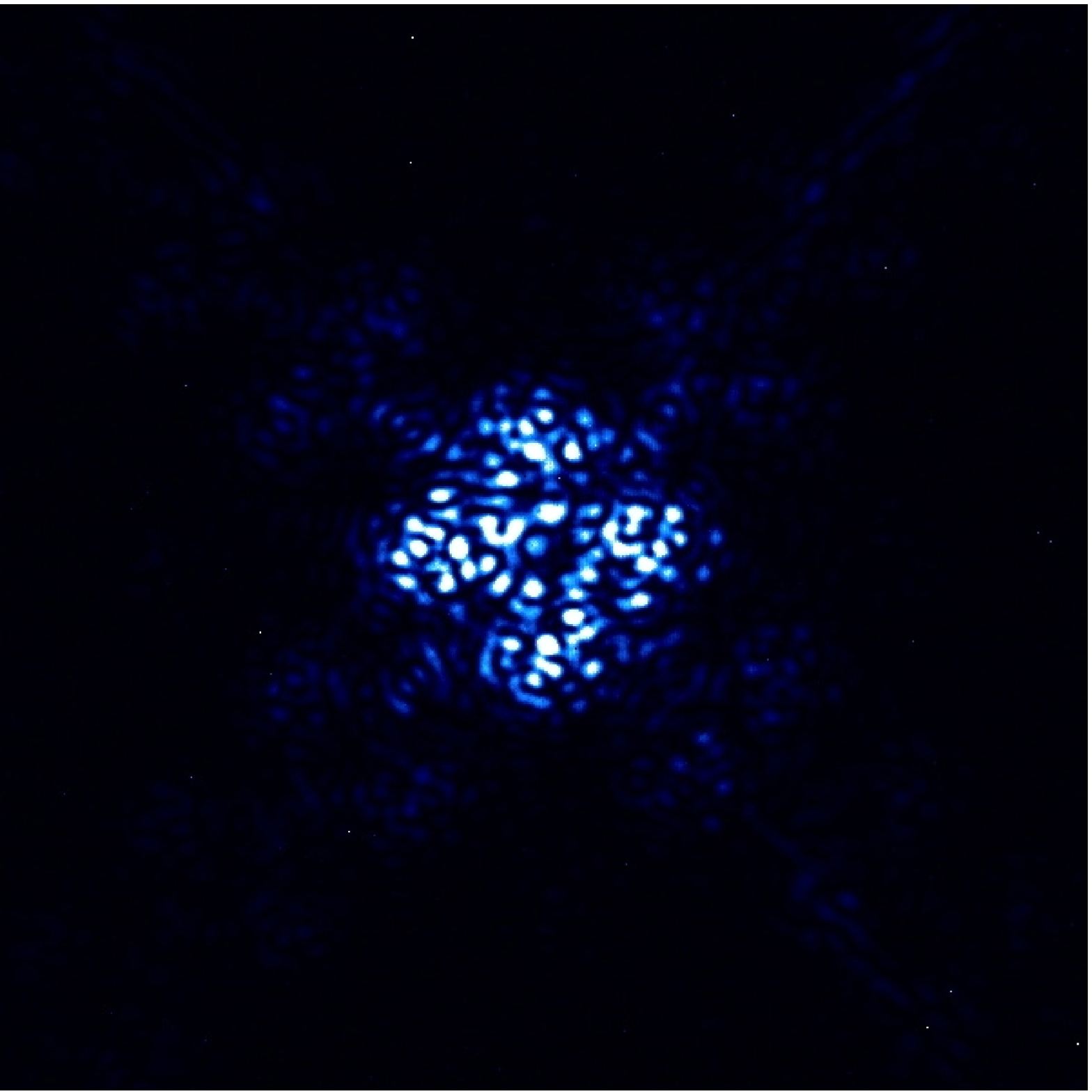}
\includegraphics[width=4.6cm]{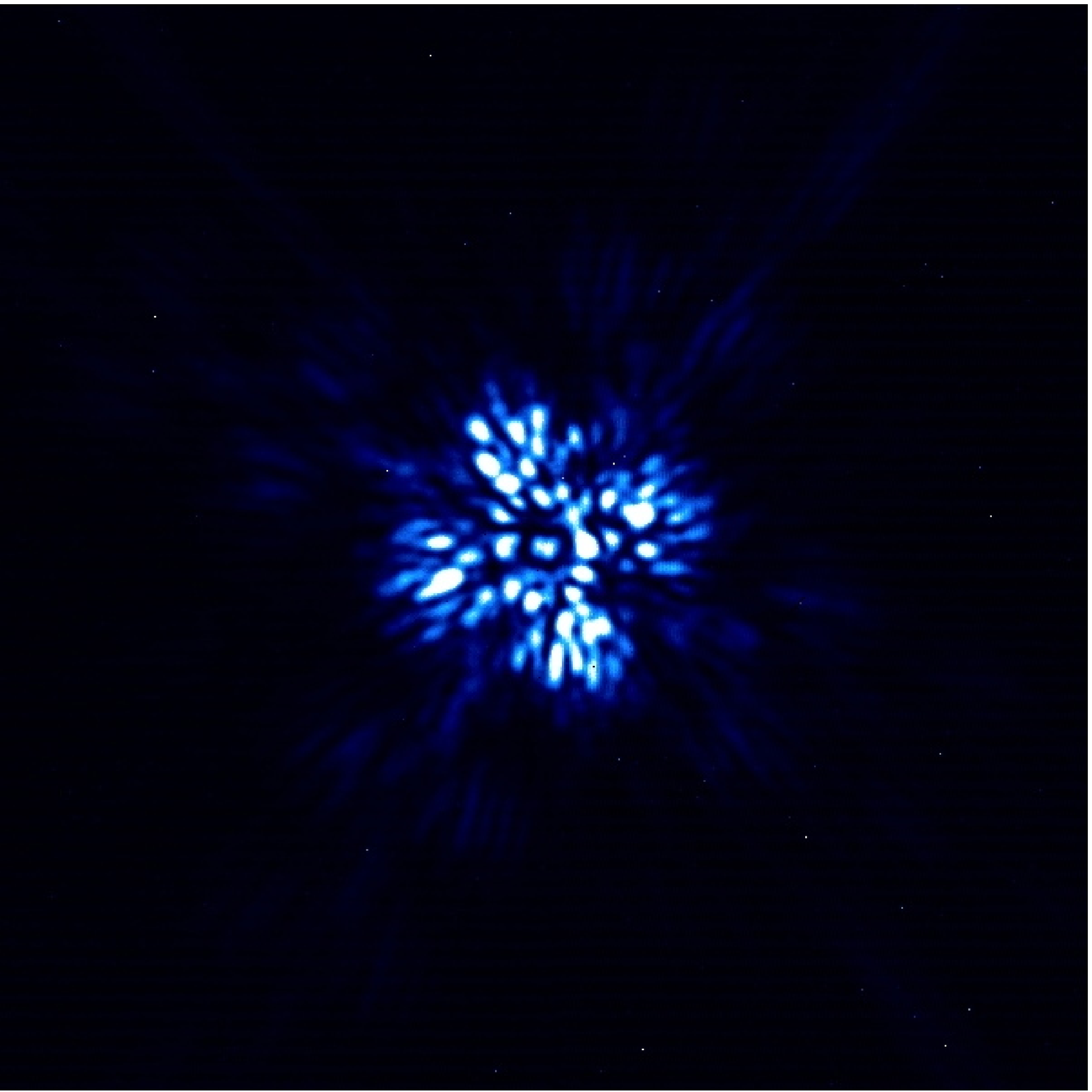}
\end{center}
\caption{Left: Shadowgraph inspection ($\times 50$) of pupil masks 2 to 6 (top row to bottom row), middle: infrared coronagraphic images ($\Delta \lambda/\lambda =1.4\%$), and on the right: infrared coronagraphic images ($\Delta \lambda/\lambda =20\%$). } 
\label{inspection}
\end{figure*} 
\begin{figure*}[!ht]
\begin{center}
\includegraphics[width=9cm]{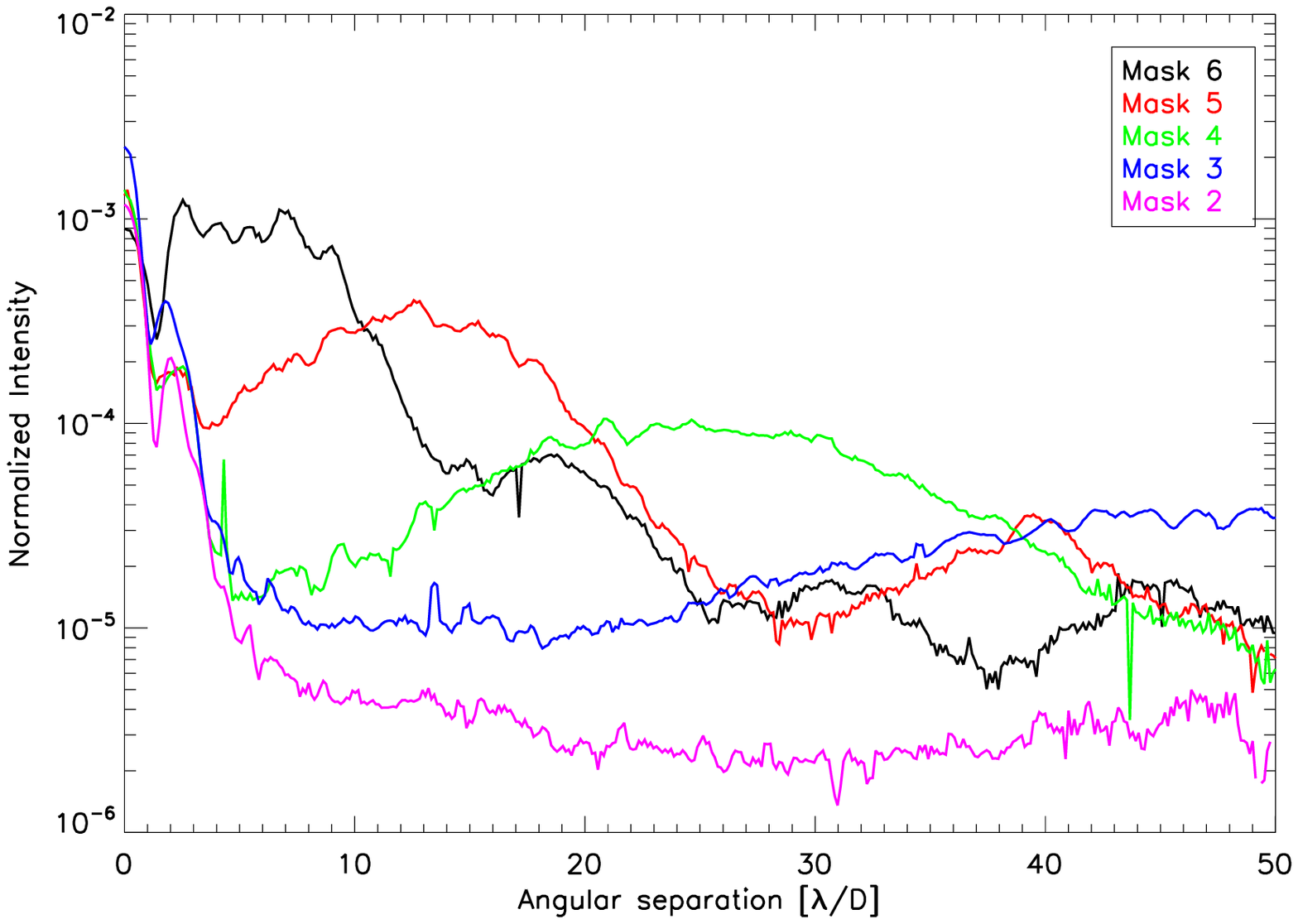}
\includegraphics[width=9cm]{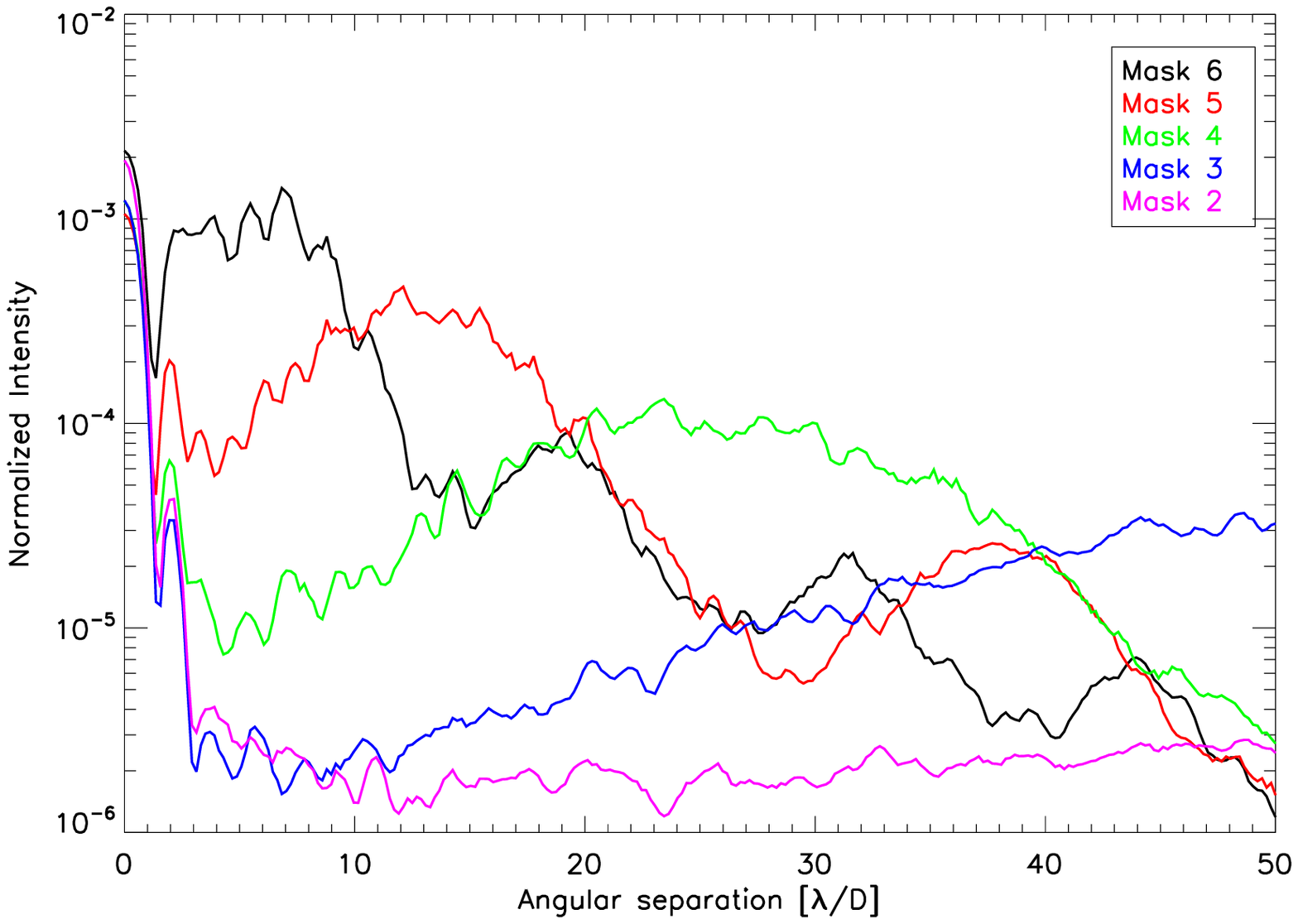}
\end{center}
\caption{Summary of coronagraphic radial profiles ($\Delta \lambda/\lambda =1.4\%$) for mask 2 to mask 6, profiles are azimuthally averaged. $Left$: recorded on the bench, $Right$: simulations assuming bench conditions.} 
\label{Results}
\end{figure*} 
\begin{center}
\begin{table*}[!ht]
\centering
\begin{tabular}{c|c|c|c|c|c|c}
\hline \hline 
Prototype & $S$ & $p$ [$\mu$m] & \multicolumn{2}{c|}{$f_g$ [$\lambda/D$]} & \multicolumn{2}{c}{$I_g$} \\
\cline{4-7} 
& & &   Theory  & Experiment   &  Theory & Experiment \\
\hline
mask 2 & 200 & 15& 108   & $107$   & $5.7\times10^{-6}$ & $6.2\times10^{-6}$ \\
mask 3 & 100 & 30 & 54   & $53$ &$2.3\times10^{-5}$ & $3.3\times10^{-5}$\\
mask 4 & 50 &  60 & 27   & 25  & $9.1\times10^{-5}$ & $9.9\times10^{-5}$\\
mask 5 & 25 & 120 & 13  & 13 & $3.6\times10^{-4}$&  $3.5\times10^{-4}$\\
mask 6 & 12.5 & 240& 7  & 7 &$1.4\times10^{-3}$& $1.1\times10^{-3}$ \\
\hline
\end{tabular}
\caption{Summary of theoretical values and laboratory measurements of the pixellation noise properties.}
\label{table1}
\end{table*}
\end{center}

\subsection{Inspection of the apodizers} 
Metrology inspection of these 5 masks has been made using a Shadowgraph ($\times 50$, see left column of Fig. \ref{inspection}). Chrome dots size has been determined to be 15, 29, 57, 119, and 240 $\mu$m $\pm 1\mu$m for mask 2 to 6 respectively, and are therefore close to the specifications. The slight pixel under-sizing of mask 3 to 5 with respect to the specifications has a negligible impact on their high-frequency noise properties.
Unlike mask 1, mask 2 to 6 were not numerically pre-compensated to avoid an increase of transmission -- as a result of a reduction of the metal dots during the wet-ech lithography process -- since dot size was less critical than for mask 1.
This pre-compensation in the digital design corrects for edges effect on the light-blocking metal dots resulting from the isotropic wet etching process (see Paper I for more details), by estimating the feature size that would be obtain after fabrication.
Pre-compensation is mainly relevant when dots size are very small, since fabrication errors gets more important.

The spatially-resolved transmissions of each apodizer has been measured.
An iris in the far field has been used to obtain the low-frequency component of each mask to verify the global shape (i.e the symmetry). Accuracy of the profile is about 3-5$\%$ in the near-IR (achromaticity has been demonstrated with mask 1 in Paper I in J and H-bands). 
The evaluation of the impact of their high-frequency contents at the coronagraphic image level is precisely the objective of this paper.

\section{Results and discussion}
Coronagraphic images recorded on the bench using masks 2 to 6 are presented in Fig. \ref{inspection} (central column: $\Delta \lambda/\lambda =1.4\%$, right column: $\Delta \lambda/\lambda =20\%$).
Speckles are clearly visible as well as speckle elongation when a broadband filter is used (Fig. \ref{inspection}, right column).
Qualitatively, the speckle halo behavior is as a function of $S$: when reducing the pixel size (from mask 6 to mask 2), the first order diffraction halo moves further away. 
When the halo is far away enough from the central core of the PSF, an usable field of view cleaned of speckles appears and reveals the residual diffraction from the pupil (spider vane diffraction spikes). 

\begin{figure}[!ht]
\begin{center}
\includegraphics[width=6.0cm]{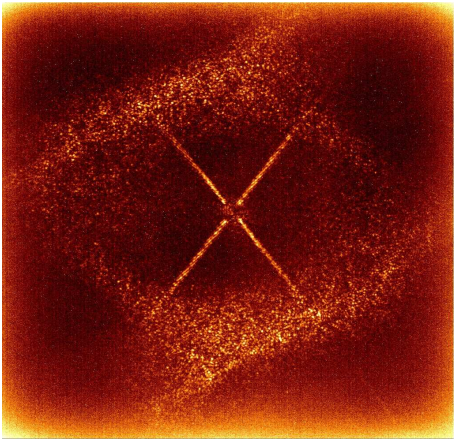}
\end{center}
\caption{Coronagraphic image using mask 2, recorded with a visible mini-camera (using a 633nm laser).} 
\label{m15laser}
\end{figure}

Coronagraphic profiles obtained with each mask are presented in Fig. \ref{Results} (left), along with simulations (Fig. \ref{Results}, right). Simulations assumed perfect microdots apodizers and bench conditions (VLT-like pupil, same bandwidth and similar Strehl ratio).
In Table \ref{table1} we compare the intensity and localization of the first order diffraction halo measured and predicted by Eq. \ref{principalF} and Eq.Ê\ref{Intensity}. The intensity has been measured on the halo peak.
All the tests performed with these 5 new masks confirmed Eq. \ref{principalF} and Eq. \ref{Intensity}.
Mask 6 and 5 (black and red curves) revealed several orders of diffraction peaks broadened by speckles, that are angularly separated by $\sim$S (in $\lambda/D$ unit), and localized at $S/2$, $3S/2$, $5S/2$ and so...
In every cases, each speckle halo has an extent in the order of $S$ in $\lambda/D$.
For Mask 2, the speckle halo was only partially imaged on the detector (limited by the detector field of view, $\sim150\times150\lambda/D$) and locally estimated at $95\lambda/D$, its averaged position in the field has been evaluated using a 633 nm laser and a visible mini-camera (see Fig. \ref{m15laser}, allowing a 360$\times$360$\lambda/D$ field of view). In such a case, the Lyot focal mask appears bigger ($\sim$$12\lambda/D$) but does not impact the halo position estimated at 108$\lambda/D$. The relative intensity of the halo meets predictions (measured locally -- in H-band --, while for previous masks, an azimuthal average has been performed).

We carried out the same test with a broadband filter in H ($\Delta \lambda/\lambda =20\%$), and we did not observe any modification of the behavior (Fig. \ref{inspection}, right column). Comparison with simulated coronagraphic profiles (Fig. \ref{Results}, right) presents a slight discrepancy, mainly for masks 2 and 3, at small angular distance without impacting the halo intensity and position. This discrepancy is within the error loss of the profile, apodizer alignment, and bench quality as discussed in Paper I.

The directionally of the noise clearly seen in Fig. \ref{inspection} (central and right columns) and Fig. \ref{m15laser} is a result of the diffusion error algorithm used to distribute the dots (left column of Fig. \ref{inspection}). Therefore, from the rise of the speckle halo, better performance are reachable in some directions. 

We further probe the far-field in the coronagraphic image when using mask 1 in visible and infrared wavelengths, but as in paper I, no noise has been revealed during the experiment.
Mask 1 was actually specified for coronagraphic application (the scaling was specified to $S = 500$ and ended up to $S=666$, see paper I), while masks 2 to 6 were precisely designed to evidence and analyze the signature of the dots in the coronagraphic image. As a result of its specification, the noise introduced by mask 1 was at such a high frequency, and low intensity, that it was not measurable on our bench.

Theoretical predictions are therefore confirmed. The simplified model used for order-of-magnitude estimation of the pixellation noise intensity in coronagraphic image is representative of the APLC situation.
Considering the validity of Eq. \ref{principalF} and Eq. \ref{Intensity}, presented in Fig. \ref{pixelsize}, we can therefore properly design microdots apodizer (i.e select pixel size) for any coronagraph concept featuring amplitude pupil apodization.
The selection of the pixel size must be defined by pushing the first order diffraction halo out of the field of interest (Eq. \ref{principalF}) and by reducing its intensity (Eq. \ref{Intensity} ) to avoid any limitations imposed even by the rise to the speckle halo. The apodizer amplitude transmission ($g$) as well as the sampling factor ($S$) drive this choice.

Ideally, going to very small pixel size improves the accuracy of the profile transmission (i.e. sampling problem).
In practice, getting good accuracy becomes more difficult because fabrication errors gets more important as the pixel size decreases, therefore an increase of the pre-compensation algorithm accuracy is required. Besides, when the pixel size is comparable to the wavelength of light, the transmission is affected by plasmons \citep{RCWA1, RCWA2}. 
Therefore, we note that in such sub-wavelength regime, a RCWA (Rigorous Coupled Wave Analysis) analysis would be mandatory for a more refined analysis of the dependency of optimal pixel size on wavelength. 

\section{Conclusion}
We reported on additional development and laboratory experiments of microdots apodizers for the Apodized Pupil Lyot Coronagraph. 
Testing different pixel size configurations allowed to confirm theoretical predictions of paper I. We demonstrated agreement between laboratory measurements and theoretical models.
The predicated coronagraphic PSF of APLC using microdots apodizer is confirmed by experiment, and any coronagraph that feature amplitude pupil apodization can be properly designed following the prediction of Eq. \ref{principalF} and Eq. \ref{Intensity}. Furthermore, the microdots technique will be the baseline approach for the apodizer of the Apodized Pupil Lyot Coronagraph for EPICS \citep{EPICS} as well as for GPI \citep{2006SPIE.6272E..18M}.

In addition, we are currently extending the technique to the manufacture of Band-Limited masks \citep{2002ApJ...570..900K}. Results of this development will be presented in a forthcoming paper.

\label{conclu}

\begin{acknowledgements}
This activity is supported by the European Community under its Framework Programme 6, ELT Design Study, Contract No. 011863.
\end{acknowledgements}
\nocite{*}
\bibliography{MyBiblio2}
\end{document}